\documentclass[a4paper,11pt]{article}

\usepackage{amsmath,amsfonts,graphicx,tikz}
\usepackage{hyperref}

\def\be{\begin{equation}}
\def\ee{\end{equation}}
\def\bea{\begin{eqnarray}}
\def\eea{\end{eqnarray}}

\newcommand{\Urm}{\mathrm{U}}

\newcommand{\SL}{\mathrm{SL}}

\newcommand{\Wcal}{\mathcal{W}}
\newcommand{\Ncal}{\mathcal{N}}
\newcommand{\Dcal}{\mathcal{D}}

\newcommand{\Lcal}{\mathcal{L}}
\newcommand{\Qcal}{\mathcal{Q}}
\newcommand{\Zset}{\mathbb{Z}}
\newcommand{\Qset}{\mathbb{Q}}
\newcommand{\Hset}{\mathbb{H}}

\newcommand{\Cset}{{\,\,{{{^{_{\pmb{\mid}}}}\kern-.47em{\mathrm C}}}}}

\newcommand{\ket}[1]{\left|#1\right.\rangle}

\newcommand{\diff}{\mathrm{d}}
\newcommand{\ra}{\rightarrow}
\newcommand{\half}{\frac12}

\newcommand{\psibar}{\overline{\psi}}

\newcommand{\bbar}{\bar{b}}
\newcommand{\tauh}{\hat{\tau}}
\newcommand{\qh}{\hat{q}}
\newcommand{\qt}{\tilde{q}}
\newcommand{\p}{\partial}

\begin{document}

\numberwithin{equation}{section}

\mbox{}

\vspace{40pt}

\begin{center}

 {\Large \bf Free fermions, neutrality and modular transformations}\\
\vspace{43pt}

{\large {\mbox{{\bf Mbavhalelo Mulokwe$\,{}^a$} \hspace{.2cm} and \hspace{.2cm} {\bf Konstantinos Zoubos$\,{}^{a,b}$}}}}%

\mbox{}
 
${}^a$ National Institute for Theoretical and Computational Sciences, School of Physics
and Mandelstam Institute for Theoretical Physics, University of the Witwatersrand,
Johannesburg, Wits 2050, South Africa

\vspace{.5cm}

${}^b$ Department of Physics, University of Pretoria\\
Private Bag X20, Hatfield 0028, South Africa

\mbox{}

Email: mulokwe.mbavhalelo@gmail.com, kzoubos@up.ac.za

\vspace{40pt}

\abstract{ With a view towards higher-spin applications, we study the partition function of a free complex fermion in 2d CFT, restricted to the neutral (zero fermion number) sector. This restriction leads to a partial theta function with a combinatoric interpretation in terms of Dyson's crank of a partition. More crucially, this partition function can be expressed in terms of a $q$-hypergeometric function with quantum modular properties. This allows us to find its high-temperature asymptotics, including subleading terms which agree with, but also go beyond, what one obtains by imposing neutrality thermodynamically through a chemical potential. We evaluate the asymptotic density of states for this neutral partition function, including the first few subleading terms. Our results should be extendable to more fermions, as well as to higher-spin chemical potentials, which would be of relevance to the higher-spin/minimal model correspondence.}

\end{center}

\vspace{.3cm}

\large

\noindent

\normalsize

\noindent

\vspace{3.6cm}

\vspace{0.5cm}

\setcounter{page}{0}
\thispagestyle{empty}
\newpage

\section{Introduction}

Functions with well-defined transformation properties under the modular group $\SL(2,\Zset)$ underlie many beautiful results in mathematics, and also enter in many physical applications. The best-known such functions are the modular forms, which are invariant under modular transformations up to a proportionality factor:
\be
\phi\left(\frac{a\tau+b}{c\tau+d}\right)=\epsilon\cdot (c\tau+d)^k \phi(\tau)
\ee
where $k$ (integer or half-integer) is called the weight, $a,b,c,d$ are integers, $\epsilon$ a complex number that depends on the $\SL(2,\Zset)$ element, and $\tau$ is a complex number taken to live on the upper-half plane $\Hset^+$. In recent years the class of modular forms has been extended significantly to functions, such as quasi-modular and quantum modular forms, for which the difference between the function and its modular transformation is not zero but exhibits ``anomalous'' behaviour, which can still have well-defined, desirable features. This has led to intense study of such more exotic modular objects and a growing appreciation of their physical relevance (see e.g. the lectures \cite{DHoker:2022dxx} for an introduction from a physics perspective, and references to the mathematical literature).  

One of the many applications of modular forms in physics has been in the study of partition functions in 3d gravity. Here the main question is to calculate the entropy of a black hole solution by counting its microstates. In settings with AdS boundary conditions, it is known that the asymptotic symmetry group is a Virasoro algebra with a central charge which depends on the solution \cite{Brown:1986nw}. Cardy's result \cite{Cardy:1986ie} is that the leading behaviour of the asymptotic density of states depends purely on the central charge of the theory, and can thus be computed without detailed knowledge of the specific dual conformal field theory (CFT). A crucial step in  Cardy's calculation involves passing from low-temperature to high-temperature behaviour, which corresponds to a modular transformation from $\tau\ra i\infty$ to $\tau\ra i0^+$ (where $1/T=\beta=2\pi (-i \tau)$). For reviews of these aspects, and in particular the role of modular transformations, we refer the reader to \cite{Carlip:1998qw,Carlip:2005zn, Murthy:2023mbc}. 

Of course, regardless of possible bulk duals, modular transformations also play a crucial role in 2d CFT as, defined on a torus, consistency requires the partition function to be modular invariant. Therefore, partition functions corresponding to CFT models such as free bosons or fermions, or minimal models, are built up from modular forms \cite{DiFrancesco}. It should be noted that to achieve modular invariance one needs to add contributions from both left- and right- movers, and, in the case of fermions, sum over contributions from sectors with different periodicities on the torus.

In this work, motivated by the higher-spin/$\Wcal$-algebra correspondence which we will briefly review in the next section, we will consider the partition function of complex free fermions in the Ramond (integer-moded) sector, where we impose the constraint of exact neutrality on the partition function. In other words, we only count states with zero fermion number. For simplicity we will focus on a single fermion (so the central charge $c=1$) and only on left-movers. We will find that, unlike the full (charged) partition function, which has standard modular behaviour, this restricted partition function is not a modular form but a partial theta function. Therefore, we do not have the full power of modularity at our disposal. However, by making use of recent work \cite{Folsometal13} which identifies this function with a quantum modular form \cite{ZagierQuantum}, we will show that it is still possible to perform a high-temperature expansion and extract the asymptotic density of states, and thus a Cardy-type formula.

In physics, quantum modular forms have already appeared in the context of indices for $3d$, $\Ncal=2$ theories, see \cite{Cheng:2018vpl} for a discussion and several examples of partial theta functions which are quantum modular forms. Partial/false theta series, understood as quantum modular forms, also appear in the context of $\Wcal$-algebra characters \cite{BringmannMilas15}. In that work, the authors also evaluated the restriction of a Jacobi form to the constant in $\zeta$ part (corresponding to restricting to the neutral sector in our case) and produced a quantum modular form. Our work can be thought of as a demonstration that similar quantum modular objects can arise in even simpler, free-field settings.

The outline of this note is as follows: After briefly discussing the overall motivation, coming from the higher-spin/minimal-model correspondence, in section \ref{Neutral} we will review the partition functions of charged and neutral complex fermions and, in section \ref{Hypergeom} express the neutral partition function for $D=1$ in terms of a $q$-hypergeometric function. This will allow us to exploit a modular formula from \cite{Folsometal13} to find the subleading corrections to the neutral partition function in section \ref{Modular}. In section \ref{Canonical} we work thermodynamically via saddle point integration and find agreement, within the applicability of the approximation, with the results from modularity. In section \ref{Asymptotic} we find the asymptotic density of states, including the subleading corrections, and we conclude with some potential future directions. Appendix \ref{appendixA} contains some definitions and useful expressions, appendix \ref{appendixB} contains our conventions for free-fermion charges, while appendix \ref{appendixC} treats the case $D=2$ to illustrate some of the expected features of the general case.

\section{Motivation: Higher spin/$\Wcal$-algebra correspondence} \label{Motivation}

\begin{table}[t]
\begin{tabular}{c| rrrrrrrrrrrr} \hline \phantom{\Large$A^A$} \!\!\!\!\!\!\!\!State& 1 &$b_{-1}$ &$\bbar_{-1}$ & $b_{-2}$ & $b_{-1}\bbar_{-1}$& $\bbar_{-2}$ &$ b_{-3}$ &$ b_{-2}b_{-1}$ & $b_{-2}\bbar_{-1}$&$ b_{-1}\bbar_{-2}$ &$ \bbar_{-2}\bbar_{-1}$ &$ \bbar_{-3}$ \\ \hline
  $\Qcal$ &0 &1 & $-1$ & 1 &0 &$-1$ &1 &2 &0 & 0&$-2$ &$-1$\\ 
  $\Lcal$ &0 & 1 &1 &2&2&2&3&3&3&3&3&3 \\
  $\Wcal$ &0 &1 &$-1$ & 4&0 &$-4$ & 9& 5&3&$-3$ &$-5$&$-9$\\ \hline
\end{tabular}
\caption{The first few states, expressed as modes acting on $\ket{0}$, contributing to (\ref{HSPartition}) for $D=1$.} \label{states}
\end{table}

A very fruitful setting for studying the AdS/CFT correspondence is the duality between higher-spin gravity on $AdS_3$ and 2d minimal models with $\Wcal$ algebra symmetry \cite{Gaberdiel:2010pz}. The duality is based on the asymptotic symmetry group of Vasiliev's higher spin theory \cite{Vasiliev:2003ev} being $\Wcal_\infty[\lambda]$ \cite{Henneaux:2010xg,Campoleoni:2010zq}, with $\lambda$ a tunable parameter. One of the most successful tests of this duality is the matching of the spectrum \cite{Gaberdiel:2011zw}. In particular, \cite{Kraus:2011ds} studied the thermodynamics of black holes with a spin-3 chemical potential and compared it to the result obtained from the CFT side. Although the duality should apply for more general CFT's, there are special points in parameter space where the CFT computation is very straightforward, and here we will focus on the case of a free fermion CFT, where the Vasiliev parameter $\lambda=0$. For this case, it is straightforward to compute the partition function \cite{Kraus:2011ds} to obtain, in the conventions of appendix \ref{appendixB}:
\be \label{HSPartition}
 Z(u,\tau,\alpha)=\mathrm{Tr}\left(e^{2\pi i(\tau\Lcal+\alpha\Wcal+ u \Qcal)}\right)=\left(\prod_{m=1}^\infty \left(1+q^mw^{m^2}\zeta\right)\left(1+q^mw^{-m^2}/\zeta\right)\right)^D
\ee
where $q=e^{2\pi i \tau}$, $w=e^{2\pi i \alpha}$, with $\alpha$ being the spin-3 chemical potential, and $\zeta=e^{2\pi i u}$, with $u$ the spin-1 charge. $D$ is the number of complex fermions, i.e. the central charge of the CFT, which on the bulk side is identified with $3\ell/(2G_N)$, with $\ell$ the AdS radius and $G_N$ Newton's constant. Notice the integer moding of the fermions as we require periodicity on the cylinder, i.e. we are in the Ramond sector, which is relevant for the BTZ black hole \cite{Coussaert:1993jp} and by extension its higher-spin cousins. Here and in the following, we are only focusing on the left-movers. 

However, a well-known complication is that the free-fermion system contains a spin-1 current and thus leads to a $\Wcal_{1+\infty}$ algebra \cite{Pope:1990kc}. This current needs to be eliminated in order to obtain the required  $\Wcal_\infty[0]$ algebra. In other words, we need to impose a neutrality condition on (\ref{HSPartition}) to be able to match with the bulk theory. In \cite{Kraus:2011ds}, this was done in the high-temperature limit $\tau\ra 0$, where one can convert the sum to an integral and differentiate with respect to $u$. This procedure, in a double-scaling limit where also $\alpha\ra0$ such that $\alpha/\tau^2$ is constant and small, led to exact agreement with the bulk entropy.\footnote{This is most immediate in the ``holomorphic'' prescription of \cite{Gutperle:2011kf}. In order to match with a canonical definition of the entropy some care is needed in the bulk/boundary identification, see \cite{Compere:2013nba} for a discussion and references.} For further work and extensions we refer to \cite{Beccaria:2013dua, Beccaria:2013gaa}.

At the level of the algebra, a consistent way to remove the $\Urm(1)$ current (at the cost of making the algebra non-linear) is to perform an order-by-order modification of the generators of $\Wcal_{1+\infty}$, such that their OPE's with the $\mathrm{U}(1)$ current vanish, allowing to take a coset by the $U(1)$ symmetry \cite{Gaberdiel:2013jpa}.\footnote{A different prescription for performing this truncation was proposed in \cite{Pope:1990be}.}  This approach has been used e.g. in \cite{Datta:2014ska} in the calculation of the entanglement entropy in the spin-3 case. As our interest is in microscopic state counting, and exploring subleading corrections (at high temperature), we will instead ask how the neutrality condition can be imposed as an exact condition on the states.

Leaving aside the neutrality constraint for the moment, it is natural to ask whether the above grand canonical partition function can be written in terms of functions with known modular properties. This would provide a link between the low and high temperature limits of the theory, similar to the standard CFT partition functions, and would potentially allow for more precise checks of the duality. (It should be emphasised that many of the above checks largely rely on the $\Wcal_\infty[\lambda]$ symmetry structure being the same, and cannot pinpoint a specific microscopic realisation, for which one would need to look at subleading terms). As already alluded to in \cite{Gaberdiel:2012yb}, a natural candidate with known modular properties is the generalised Jacobi function introduced in \cite{KanekoZagier}, defined through its (formal) $q$-series as
\be \label{HKZ}
\Theta(\zeta,q,w)=\prod_{n>0}(1-q^n)\prod_{n>0,~ \text{n odd}}^\infty \left(1-q^{n/2}w^{n^2/8}\zeta\right) \left(1-q^{n/2}w^{-n^2/8}/\zeta\right)\;.
\ee
It is straightforward to express (\ref{HSPartition}) in terms of these functions, to obtain\footnote{For a detailed derivation of the first formula, the reader is referred to the MSc dissertation \cite{LukhoziMSc}. The relation to the second expression is just the higher-spin version of the equality $\theta_2(u|2\tau)\theta_3(u|2\tau)/\eta(2\tau)^2=\theta_2(u|\tau)/\eta(\tau)$.}
\be
Z(u,\tau,\alpha)=\left(w^{1/6}q^{1/6}\frac{\zeta}{1+\zeta} \frac{\Theta(-\zeta,q^2,w^8)\; \Theta(-w q\zeta,w^{4}q^2,w^8)}{\eta(2\tau)^2}\right)^D\;,
\ee
or equivalently
\be \label{Zuta}
Z(u,\tau,\alpha)=\left(w^{1/24}q^{1/24}\frac{\zeta}{1+\zeta} \frac{\Theta(-w^{\frac{1}{4}} q^\half\zeta, w q,w^2)}{\eta(\tau)}\right)^D\;.
\ee
Although these expressions reproduce (\ref{HSPartition}), recall that this is not the partition function that we actually want, as it does not take the neutrality condition into account. In the thermodynamic limit, one can of course always impose neutrality by passing to the canonical ensemble. However, that by itself does not provide much insight into the modular properties of the neutral partition function. Ideally, then, one would like to impose neutrality as an exact condition, and not just in the high-temperature limit. So we are interested in understanding the $\zeta^0$ part of the above expression.

Now, one of the main results of \cite{KanekoZagier} was that the $\zeta^0$ component of $\Theta(\zeta,q,w)$ can be expanded in powers of $w^{2n}$, the coefficients of which are quasi-modular forms of degree $6n$. This result played an important role in the counting of maps from genus-$g$ curves to genus 1 \cite{Dijkgraaf94}, where in that case $n=g-1$. In that case the underlying free-fermionic description \cite{Douglas:1993wy} is half-integer moded, so the specific $\Theta$ function appearing is ``$\theta_3$-like'' and in particular its expression is a sum is over integer powers (see appendix \ref{appendixA}). The required $\Theta$ in our case (\ref{Zuta}) is ``$\theta_2$-like'' with half-integer powers in the sum, so (in either of the forms above) extracting the $\zeta^0$ terms is slightly more delicate. We will get a flavour of this difference in the next section, for the $\alpha=0$ case, and we will see that it will lead to an object with quantum- instead of quasi- modular properties.

Having explained our motivation, we will not discuss the higher-spin setting further in this work, and will instead focus on the much more modest goal of imposing neutrality in the $\alpha=0$ case, and only for one fermion, i.e. $D=1$. (The $D=2$ case is treated in appendix \ref{appendixC}).

\section{Neutrality, Dyson's crank, and partial theta functions} \label{Neutral}

Let us now (and for the rest of this paper) consider the free CFT consisting of one complex fermion ($D=1$ above), with no higher-spin chemical potential turned on. In this case (\ref{HSPartition}) expands as
\be \label{Zfull}
\begin{split}
Z(u,\tau)&=1+(\zeta+\zeta^{-1})q+(\zeta+1+\zeta^{-1})q^2+(\zeta^2+\zeta+2+\zeta^{-1}+\zeta^{-2})q^3+\cdots\\
&=\prod_{m=1}^\infty (1+q^m\zeta)(1+q^m/\zeta)\;,
\end{split}
\ee
which can be expressed as
\be \label{Zfulltheta}
Z(u,\tau)=q^{-\frac1{12}}\frac{\theta_2(u|\tau)}{2\cos(\pi u)\eta(\tau)} \;.
\ee
One is often interested in counting all states in this theory, regardless of their charge. In that case one can set $u=0$ and obtain the well-known series \cite{fullpartition}:
\be
Z_{\text{charged}}(\tau)=Z(u=0,\tau)=1+2q+3q^2+6q^3+9q^4+14 q^5+22q^6+32q^7+46q^8+\cdots
\ee
To see the high-temperature behaviour of $Z_\text{charged}$ we need to study it as $\tau\ra i0^+$, where $q\ra 1$ and the $q$-expansion is not very helpful. As is standard, we apply modular transformations (see appendix \ref{appendixA}) to rewrite (\ref{Zfulltheta}) as
\be \label{Zutmodular}
Z(u,\tau)=q^{-\frac{1}{12}}\frac{e^{\pi i u^2\tauh}}{2\cos(\pi u)}\frac{\theta_4(-u\tauh|\tauh)}{\eta(\tauh)} \;,
\ee
where $\tauh=-1/\tau$ and thus $\tauh\ra i\infty$, so that $\qh=e^{2\pi i \tauh}\ll1$. Setting also $u=0$, the leading behaviour comes from the $\eta(\tauh)$ which gives a factor of $e^{\frac{-\pi i\tauh}{12}}=e^{\frac{\pi}{12\tau}}$:
\be \label{Zchargedasympt}
Z_{\text{charged}}(\tau) \sim \frac{e^{\frac{\pi i}{12\tau}-\frac{\pi i \tau}{6}} }2\;.
\ee
So we get the usual leading behaviour (giving the Cardy asymptotics for $c=1$) as $\tau\ra i0^+$:
  \be
  \ln Z_{\text{charged}}(\tau)=\frac{\pi i}{12\tau}+\cdots \;\; \text{or equivalently}\;\; -\ln q\cdot \ln Z_{\text{charged}}=\frac{\pi^2}{6}+\cdots
  \ee
This agrees with the $\alpha^0$ term in the $\lambda=0$ expansion of \cite{Kraus:2011ds}, even though we haven't yet imposed any neutrality requirement. Clearly, the difference between the charged and neutral partition functions will be in the subleading  (higher-order in $\tau$) terms, as we will see explicitly below.

Now let us turn to the opposite limit, where we want to only count the neutral states. We will call this the neutral partition function $Z_{\text{neutral}}(\tau)$, and by inspection its series starts as
\be \label{Zneutexp}
Z_{\text{neutral}}(\tau)=1+q^2+2q^3+3q^4+4q^5+6q^6+8q^7+12q^8+16q^9+23 q^{10}+30 q^{11}+42 q^{12}+\cdots
\ee
This is also a well-known series in the OEIS \cite{Nonnegativecrank}. Before we discuss its combinatoric interpretation, let us find its general series expansion. We just need to isolate the $\zeta^0$ component of (\ref{Zfulltheta}). One simply uses the definition of $\theta_2(u|\tau)$ as a sum to write
\be  \label{Zrestricted}
\begin{split}
  Z(u,\tau)&=\frac{q^{-\frac1{12}}}{\eta(\tau)}\frac1{\zeta^{\half}+\zeta^{-\half}}\sum_{r\in \Zset+1/2}\zeta^r q^{r^2/2}\\
  &=\frac{q^{-\frac1{12}}}{\eta(\tau)}\frac1{\zeta^{\half}+\zeta^{-\half}}\sum_{k>0,\; k \; \text{odd}}(\zeta^{k/2}+\zeta^{-k/2})q^{k^2/8}\\
  &=\frac{q^{-\frac1{12}}}{\eta(\tau)}\sum_{k>0,\; k\; \text{odd}} \left(\sum_{\ell=-(r-1)/2}^{\ell=(r-1)/2}(-1)^{\frac{k-1}2-\ell}\zeta^\ell\right)q^{k^2/8}\;,
\end{split}
\ee
where the sum over $\ell$ comes from dividing $\zeta^{k/2}+\zeta^{-k/2}$ by $\zeta^\half+\zeta^{-\half}$.  The $\zeta^0$ components in the last line of (\ref{Zrestricted}) are those where $\ell=0$, which leads to an alternating $(-1)^{\frac{k-1}2}$ factor for each $r$. So
\be \label{restrict}
Z_{\text{neutral}}(\tau)=\oint \frac{\diff\zeta}{2\pi i \zeta} Z(u,\tau)=\frac{q^{-\frac1{12}}}{\eta(\tau)}\sum_{k>0,~k~ \text{odd}}(-1)^{\frac{k-1}2}q^{k^2/8}=
\frac{q^{\frac1{24}}}{\eta(\tau)}\sum_{n\geq 0}(-1)^{n}q^{\frac{n(n+1)}2}\;,
\ee
where we set $k=2n+1$ in the last step. 
This series has recently been shown to be a generating function of Dyson's non-negative crank partition function, see \cite{Uncu2018,AndrewsNewman20}, where it is given in the equivalent form
\be \label{ZUncu}
Z_{\text{crank}\geq 0}(q)=\frac{1}{(q;q)_{\infty}}\sum_{n\geq 0}(-1)^n q^{\binom{n+1}{2}}\;,
\ee
with $(q;q)_{n}$ the $q$-Pochhammer symbol (see appendix \ref{appendixA}).

To better understand the combinatorics behind this series, it is useful to recall Dyson's definitions of the rank and crank of a partition.   
In \cite{Dyson44}, Dyson defined the rank of a partition as the largest part minus the number of components of the partition. He also explained why another property, the \emph{crank} of a partition, should exist and outlined its expected properties, however without providing a definition. A definition of the crank was finally given in \cite{AndrewsGarvan88}, and it reads
as follows: If the partition does not contain 1's, it is the largest part of the partition, otherwise it is the number of parts larger than the number of 1's, minus the number of 1's. 

As an example, considering the partitions of 5, the crank of the partition $(3,2)$ is 3, while that of the partition $(3,1,1)$ has two 1's and one part larger than 2, so the crank is $1-2=-1$. For the partition $(2,2,1)$, we have one factor of 1, and there are two parts which are larger than 1, so the crank is $2-1=1$ etc. 

We see that the crank can be positive, zero, or negative. In Table 2 we list the cranks corresponding to the first few natural numbers.
\begin{center}
\begin{tabular}{c|c|c|c} $n$ & $p_i\;,\;\;i=1\ldots p(n)$ & $c(p_i)$ & $a(n)$\\ \hline 
  0 & \{\} & 1&1\\
  1 & \{1\} & -1&0 \\
  2 & \{2\},\{$1^2$\} & 2,-2&1\\
  3 & \{3\},\{2,1\},\{$1^3$\} & 3,0,-3&2\\
  4 & \{4\},\{3,1\},\{2,2\},\{2,$1^2$\},\{$1^4$\} & 4,0,2,-2,-4&3\\
  5 & \{5\},\{4,1\},\{3,2\},\{3,$1^2$\},\{2,2,1\},\{2,$1^4$\},\{$1^5$\} & 5,0,3,-1,1,-3,-5&4\\
  6 & \{6\},\{5,1\},\{4,2\},\{4,$1^2$\},\{3,3\},\{3,2,1\},\{3,$1^3$\},\{2,2,2\}, & 6,0,4,-1,3,1,-3,2, &6\\
    & \{2,2,$1^2$\},\{2,$1^4$\},\{$1^6$\} & -2,-4,-6  &
\end{tabular}
\end{center}
In the last column, we have named $a(n)$ the number of partitions of $n$ with non-negative crank. Writing the generating function
\be
Z_{\text{crank}\geq 0}(q)=\sum_{n\geq 0} a(n) q^n\;,
\ee
we can confirm agreement with (\ref{ZUncu}) to arbitrary order, and thus with $Z_{\text{neutral}}$.

To conclude, imposing neutrality as a strict condition corresponds to counting partitions with non-negative crank. It would be interesting to study whether Dyson's crank might enter in other cases where we impose a physical restriction on a partition function, and also whether other charge sectors (e.g. $Q=1$ or $|Q|\leq Q_{\text{max}}$ for some $Q_{\text{max}}$) might also lead to interesting combinatorial objects. 

Let us note that there exist other possible expressions for the non-negative crank partition function, such as \cite{Nonnegativecrank}:
\be
Z_{\text{neutral}}(\tau)=\sum_{n=0}^{\infty}\frac{q^{n(n+1)}}{(q;q)_{n}^2}\;,
\ee
however it is the alternating form (\ref{ZUncu}) that seems to arise more naturally from the restriction to the neutral sector.

Identifying $Z_{\text{neutral}}$ as a known combinatorial series does not by itself tell us much about its modular behaviour, if any. In particular, the sum in (\ref{restrict}) is a \emph{partial theta function}, i.e. it would be a true theta function if the sum extended over all the integers, and such functions (and similarly the closely related false theta functions) typically do not have modular properties. However, it turns out that this series does have well-defined asymptotic behaviour, identified already by Ramanujan. From a modern perspective, this behaviour can be explained by relating it to a simple $q$-hypergeometric function which can be shown to have modular properties by identifying it as a quantum modular form. We will discuss these aspects in the next section. 

As mentioned in the Introduction, a computation very similar to (but more general than) (\ref{Zrestricted}) was performed in \cite{BringmannMilas15}, in the context of $\Wcal$-algebra characters. In particular, the authors showed how restricting to the $\zeta^0$ term of a Jacobi theta function leads to a quantum modular form and this allowed them to compute its asymptotic behaviour. 

As an aside, the need to restrict (functions of) Jacobi forms to what we call the neutral sector arises in other contexts, most prominently the computation of the Schur index of 4d $\Ncal=2$ superconformal gauge theories, where the requirement arises due to gauge invariance. In those cases the pole structure of the relevant contour integrals is far more intricate
than in (\ref{restrict}), see \cite{Razamat:2012uv,Pan:2021mrw} for examples. The fact that the resulting functions have less standard modular properties, transforming as quasi-modular/quasi-Jacobi forms \cite{Beem:2017ooy,Pan:2021mrw,Beem:2021zvt}, is an important element of those studies.

\section{$Z_{\text{neutral}}$ as a $q$-hypergeometric function} \label{Hypergeom}

As part of their investigations of quantum modular forms, the authors of \cite{Folsometal13} studied a class of ``Rogers-Fine'' partial theta functions on $\Hset^+$ which, through their definition as $q$-hypergeometric functions, can be related to mock modular forms \cite{ZwegersThesis} on $\Hset^-$, which therefore have (vector-type) modular transformation properties. In terms of the $q$-hypergeometric series 
\be
F(\alpha,\beta,z;q)=\sum_{n=0}^\infty \frac{(\alpha q;q)_nz^n}{(\beta q;q)_n}\;,
\ee
these functions are defined as
\be
H(a,b;\tau)=q^{1/8} F(\zeta_b^{-a}q^{-1},\zeta_b^{-a},\zeta_b^{-a}q;q^2)\;,
\ee
and
\be
G(a,b;\tau)=\frac{q^{a^2/b^2}}{1-q^{a/b}} F(-q^{a/b-1},q^{a/b},-q^{a/b};q)\;,
\ee
where $\zeta_n=e^{2\pi i/n}$. Specialising to $a=1$ and $b=2$, one finds
\be
H(1,2;\tau)=\sum_{n=0}^\infty(-1)^n q^{(2n+1)^2/8}=q^{1/8}\sum_{n=0}^\infty(-1)^n q^{\half n(n+1)}\;.
\ee
and it turns out that $G(1,2;\tau)=H(1,2;2\tau)$. 

\begin{figure}
  \begin{center}
\begin{tikzpicture}[baseline={(0,0cm)}]
  \node at (1,2) {\includegraphics[width=7cm]{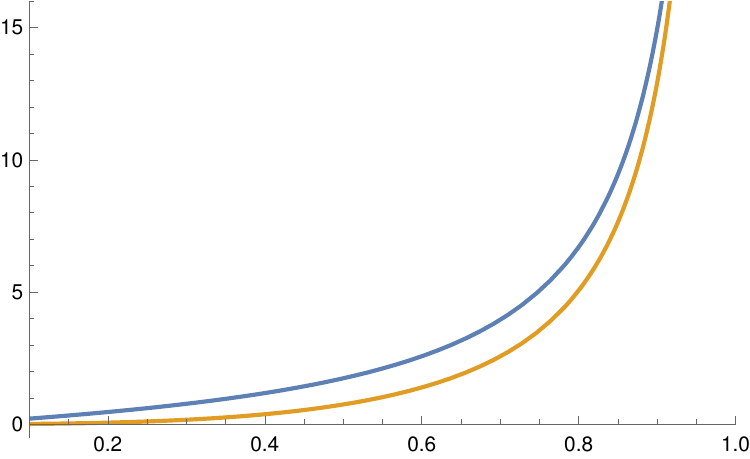}};
  \node at (2,-0.5) {$q$};
  \node at (1.5,1.7){$\scriptsize\ln Z_{\text{charged}}$};
  \node at (3.5,0.7){$\scriptsize\ln Z_{\text{neutral}}$};\node at (5.5,0) {\mbox{}};
\end{tikzpicture}
\begin{tikzpicture}[baseline={(0,0cm)}]
  \node at (1,2) {\includegraphics[width=7cm]{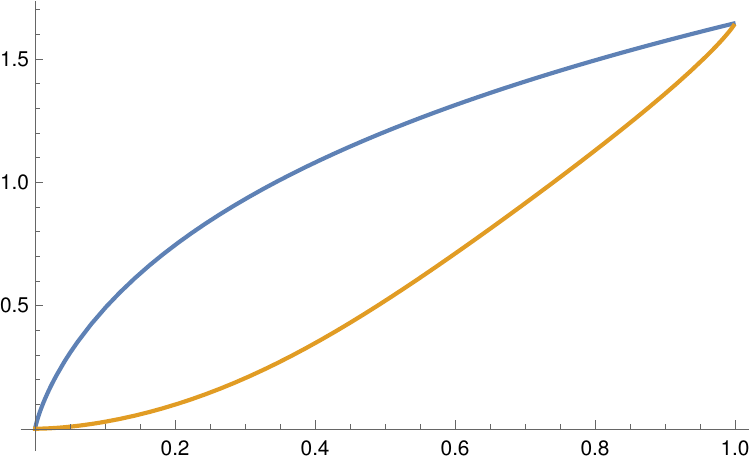}};
  \node at (2,-0.5) {$q$};
  \node at (0.45,3.5){$\scriptsize-\ln q\cdot\ln Z_{\text{charged}}$};
  \node at (2.5,1.2){$\scriptsize-\ln q\cdot \ln Z_{\text{neutral}}$};
\end{tikzpicture}

  \end{center}
  \caption{A comparison of the charged ($u=0$) partition function $Z_{\text{charged}}$ and the neutral partition function $Z_{\text{neutral}}$. On the left we plot $\ln Z_\text{charged}$ and $\ln Z_{\text{neutral}}$ while on the right we multiply them by $-\ln q$ for additional clarity. As discussed in the text, the limiting value is $\zeta(2)=\pi^2/6=1.64493\ldots$.  One sees that, as expected, the neutral partition function starts off slower than the charged one at low temperature ($q\ll 1$), but ends up having the same asymptotic behaviour at high temperature ($q\ra 1$).  We are interested in seeing this difference in the subleading behaviour analytically, by performing an expansion away from $q\ra 1$.} \label{ZfZn}
  \end{figure}
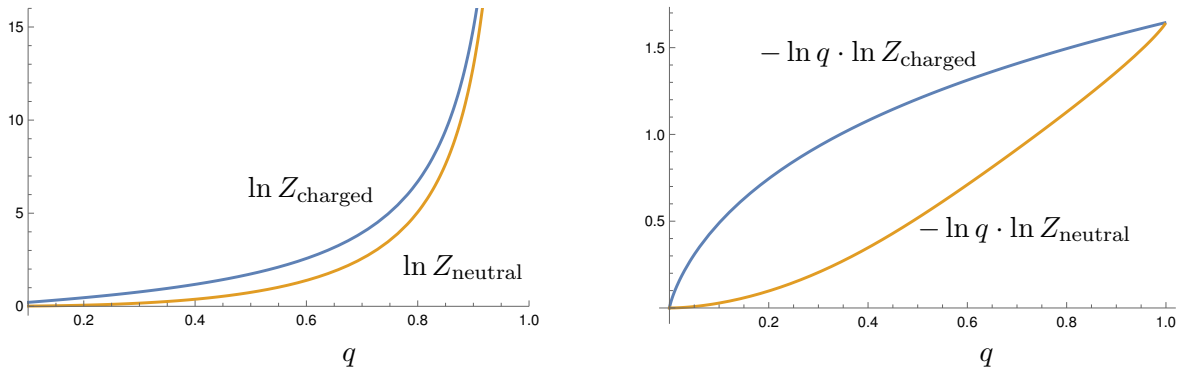

So, given (\ref{ZUncu}), we can express $Z_{\text{neutral}}$ as the $q$-hypergeometric function:
\be \label{Zneuthypergeom}
Z_{\text{neutral}}(\tau)=\frac{1}{(q;q)_{\infty}} F(-q^{-1},-1,-q;q^2)=\frac{q^{-\frac{1}{12}}}{\eta(\tau)} H(1,2;\tau)\;.
\ee
Having an exact formula allows us to evaluate $Z_{\text{neutral}}$ numerically for any required values of $\tau$, and, for instance, compare it with the charged partition function as shown in Fig. \ref{ZfZn}. We see that, as expected, the two partition functions have the same asymptotics as $q\ra 1$ but differ in their subleading behaviour. 

More crucially, being able to write a given $q$-series as a $q$-hypergeometric function goes a long way towards being able to extend it to the opposite half-plane from its original definition, see \cite{Rhoades18,Cheng:2018vpl} for more details and examples. Indeed, it was shown in \cite{Folsometal13} that the functions $H(a,b;\tau)$ inherit modular transformation properties from their lower-half-plane analogues. In the next section we will review these properties and see how we can use them to extract the high-temperature behaviour of $Z_{\text{neutral}}$ analytically.

\section{Modularity and high-temperature expansion} \label{Modular}

As we saw, restricting the free-fermion partition function to the neutral sector gave us a partial theta function. The asymptotic behaviour of such functions has been studied by Ramanujan, who for the specific case of interest obtained, for $t$ small and positive  \cite{Ramanujan2}:
\be \label{Ramanujanasympt}
\sum_{n\geq 0}(-1)^n \left(\frac{1-t}{1+t}\right)^{n(n+1)}=\half\left(1+t+t^2+2t^3+5t^4+17t^5+\cdots\right)\;.
\ee
Expressing $t=(1-q^\half)/(1+q^\half)$ and expanding $q=e^{2\pi i\tau}$ in $(-i\tau)$, we find that as $\tau\ra i0^+$,
\be
\sum_{n\geq 0}(-1)^n q^{\frac{n(n+1)}2}=\half\left(1+\frac{\pi(-i\tau)}{2}+\frac{\pi^2(-i\tau)^2}4+\frac{5\pi^3(-i\tau)^3}{24}+\cdots\right)\;,
\ee
and thus for $H(1,2;\tau)$:
\be \label{Hexpansion}
H(1,2;\tau)=q^{\frac{1}{8}}\sum_{n\geq 0}(-1)^n q^{\frac{n(n+1)}2}=\half\left(1+\frac{\pi(-i\tau)}4+\frac{5\pi^2(-i\tau)^2}{32}+\frac{61\pi^3(-i\tau)^3}{384}+\cdots\right)\;.
\ee
A proof of (\ref{Ramanujanasympt}) can be found in \cite{Berndt85} (p. 545), see also \cite{BerndtYee03,BerndtKim11,BringmannKim16,Bringmannetal17} for further discussion and extensions to more general partial theta functions. A remarkably simple way of expressing the result can be found in \cite{Bringmannetal17}, where the authors consider the following class of partial theta functions:
  \be
  F_{d,\ell}(z;\tau)=\sum_{n\geq 0}\zeta^{\ell n+d}q^{(\ell n+d)^2} \;\;,\text{where}\;\; d\in \Qset \;\;\text{and}\;\; \ell\in {\mathbb N}\;,
  \ee
  where here $\zeta=e^{2\pi i z}$ and as always $q=e^{2\pi i \tau}$. It is easy to check that for $\ell=1$, $d=1/2$ and $z=1/2$ one obtains the $G(1,2;\tau)$ function defined above, which equals $H(1,2;2\tau)$. Then (Corollary 4.6 of \cite{Bringmannetal17}) the authors show that for $\tau$ on the positive imaginary axis this function expands as
  \be \label{Fdlexp}
  F_{d,\ell}(z;\tau)=\sum_{a=0}^N\Dcal_z^{2a}\left(\frac{e^{2\pi i d z}}{1-\zeta^\ell}\right)\frac{(2\pi i \tau)^a}{a!}+O(\tau^{N+1})\;,
  \ee
  where $\Dcal_z=\frac{1}{2\pi i}\frac{\p}{\p z}$. Evaluating this expression, setting $z=1/2$ and remembering to take $\tau\ra \tau/2$, we reproduce (\ref{Hexpansion}) to any desired order $N$.

Although we now have our asymptotics, we would still like to understand them in terms of a modular transformation law, which might provide a better starting point for generalisations. It is therefore important that in \cite{Folsometal13} this series was identified as a quantum modular form. As defined in \cite{ZagierQuantum}, a quantum modular form of weight $k$ is a complex-valued function on the rationals $\Qset$, such that the difference
\be
\phi(\tau)-\epsilon^{-1} \cdot (c\tau+d)^{-k} \phi\left(\frac{a\tau+b}{c\tau+d}\right)
\ee
(with $k$ the weight) has some desired behaviour, such as analyticity. One commonly imposes a stronger requirement, that the function extends to an analytic function on $\Hset^+\cup\Hset^-$, whose asymptotics as one approaches any rational point on the real line (either from above or below) agrees with the formal power series of the function at that point. A major application of this strong version of quantum modular forms is that they allow to connect functions between the upper and lower-half planes, such that modular properties defined on the lower half-plane can filter through the rationals and extend to the upper-half plane, though only perturbatively in $\tau$. (See \cite{Cheng:2018vpl} for a treatment emphasising this crucial feature of quantum modular forms in a physical context). In such a way, series such as partial theta functions, which do not have true modular properties, if identified as quantum modular forms can be seen to have a version of a modular transformation, perturbatively on the upper-half plane.

In \cite{Folsometal13} it was shown that the $H(a,b;\tau)$ defined above, are indeed (vector-valued) strong quantum modular forms of weight $1/2$. Their transformation law for $\tau$ in $\Qset$, extended to the upper-half plane, is \cite{Folsometal13}
\be
G(a,b;-\tau)+\frac{e^{-\pi a/b}}{\sqrt{2 i \tau}} H(a,b;\frac{1}{2\tau})=-\int_0^{i\infty}\frac{(-iv)^{-3/2} T(a,b;-1/v)}{\sqrt{-i(v+\tau)}}~\diff v\;,
\ee
where $G(a,b;\tau)$ was defined above and $T(a,b;v)$ is a specific modular form of weight $3/2$. We see that the anomalous modular behaviour is expressed in terms of what is known as an Eichler integral, which in the integer-weight case is a way of expressing the polynomial anomalous dependence on $\tau$, and can be seen to play the same role in the half-integer case. In the following we will not need the general definitions but only those related to $H(1,2;\tau)$. Recall that in this case $G(1,2;\tau)=H(1,2;2\tau)$,
    and the function $T(1,2;v)$ is \cite{Folsometal13}
    \be
  T(1,2;v)=i\sum_{n=-\infty}^\infty(n+\frac14) \qt^{(n+1/4)^2}=\frac{1}{4}\sum_{k=0}^{\infty}(-1)^k (2k+1) \qt^{(2k+1)^2/16}\;,
      \ee
(here $\qt=e^{2\pi i v}$) from which we can recognise it as a $\theta_1'$ function (see appendix \ref{appendixA}), more specifically:
\be
T(1,2;v)=\frac{i}{8\pi} \theta_1'(0|v/2)= \frac{i}{4}\left(\eta(v/2)\right)^3\;,
\ee
which is indeed modular of weight 3/2. We can now rescale $\tau\ra \tau/2$ and find a transformation involving just $H$:
\be \label{Hmodular}
H(1,2;-\tau)=\frac{-i}{\sqrt{i\tau}} H(1,2;1/\tau)- \int_0^{i \infty}\frac{(-iv)^{-3/2} T(1,2;-1/v)}{\sqrt{-i(v+\tau/2)}} ~\diff v\;.
\ee
As indicated, this transformation is valid on the lower-half-plane while of course we need it on the upper-half plane, in particular we want to relate $\tau\ra i0^+$ to $\tau\ra i\infty$. However, as discussed, one of the crucial features of quantum modular forms is that one can extend such formulas to the upper-half plane, although only perturbatively in $\tau$ (see e.g. \cite{LawrenceZagier} for a discussion of this property, which inspired the later definition of quantum modular forms). We are of course interested precisely in the small $\tau$ expansion of our partition function. Taking $\tau\ra -\tau$ and expanding the denominator of the Eichler integral in (\ref{Hmodular}), we find 
  \be
  I=I_0+I_1 \frac{-i\tau}4+I_2 \frac{3(-i\tau)^2}{32}+I_3 \frac{5(-i\tau)^3}{128}+I_4 \frac{35(-i\tau)^4}{2048}+I_5\frac{63(-i\tau)^5}{8192}+\cdots
  \ee    
where we have defined
  \be
  I_k=-\frac{i}{4}\int_0^{i\infty}(-iv)^{-k-2} \left(\eta\left(-\frac1{2v}\right)\right)^3~\diff v\;.
  \ee
  Rescaling $v\ra2v$ and using the modular property of the Dedekind eta function, we can simplify these integrals to
  \be
I_k=- 2^{k-1} i \int_0^{i\infty} (-iv)^{-k-\half} \left(\eta(v)\right)^3\diff v\;.
  \ee
  There are certainly more powerful approaches to such integrals over the modular parameter, but for this simple case we find it convenient to use the identity (see \cite{Glasser2009}):
  \be
  \int_0^{\infty} f(x)~\left(\eta(ix)\right)^3\diff x=\int_0^\infty \frac{F(t)}{\cosh\sqrt{\pi t}}\diff t\;,
  \ee
  with $F(t)$ the inverse Laplace transform of $f(x)$. Identifying $v=ix$ and using \cite{AbramowitzStegun} that for $f(x)=x^{-k-1/2}$, $F(t)=2^k t^{k-1/2}/((2k-1)!!\sqrt\pi)$,  we obtain
  \be
  I_k=\frac{2^{2k-1}}{(2k\!-\!1)!! \pi^\half}\int_0^\infty \frac{t^{k-\half}}{\cosh\sqrt{\pi t}}\diff t=\frac{(2k)!}{(2k\!-\!1)!! 2^{2k+1}\pi^{k+1}}\left(\zeta\left(2k\!+\!1,\frac14\right)-\zeta\left(2k\!+\!1,\frac34\right)\right)\;.
  \ee
  The specific combination of Hurwitz $\zeta$-functions appearing here gives $\pi^{2k+1} \times$ rational numbers, which can be evaluated (with some care for the case $k=0$) to find
  \be
    I_0=\half\;\;,\;\;I_1=\frac{\pi}{2}\;\;,\;\;I_2=\frac{5\pi^2}{6}\;\;,\;\;I_3=\frac{61\pi^3}{30}\;\;,\;\;I_4=\frac{277\pi^4}{42}\;\;,\;\;
    I_5=\frac{50521\pi^5}{1890}\;\;,\ldots
  \ee
  Together with the fact that $H(1,2;-1/\tau)$ is exponentially small as $\tau\ra i0^+$ (due to the $\hat{q}^{1/8}$ factor, where $\hat{q}=e^{-2\pi i/\tau}$), we arrive at the small-$\tau$ expansion of $H(1,2;\tau$):
  \be \label{Hexpansionint}
  H(1,2;\tau)=\half+\frac{\pi}{8}(-i\tau)+\frac{3\cdot 5  \pi^2}{32\cdot 6}(-i\tau)^2+\frac{5\cdot 61 \pi^3}{128\cdot 30}(-i\tau)^3+
  \frac{35\cdot 277\pi^4}{2048\cdot 42}(-i\tau)^4+    \cdots
  \ee
which agrees with (\ref{Hexpansion}).   Although the formula (\ref{Fdlexp}) is far simpler computationally, we believe that the computation leading to (\ref{Hexpansionint}) also has its merits, as it makes it more clear which numerical factors come from the expansion of the denominator and which from the integrals, and might have more general applicability in future cases where the analogue of (\ref{Fdlexp}) might not be available. 

  We are now ready to substitute into $Z_{\text{neutral}}$, to find its small $\tau$ ($q\ra 1$) expansion (up to exponentially small terms): 
  \be \label{Znexpanded}
  Z_{\text{neutral}}(\tau)=q^{-\frac{1}{12}}\frac{\sqrt{-i\tau}}{2\eta(-1/\tau)}\left(1 +\frac{\pi(-i\tau)}{4}+\frac{5 \pi^2}{32 }(-i\tau)^2+\frac{ 61 \pi^3}{384}(-i\tau)^3+\cdots\right)\;,
  \ee
  where we also used the modular transformation $1/\eta(\tau)=\sqrt{-i\tau}/\eta(-1/\tau)$. For small $\tau$, this expansion can be seen to agree very well numerically with the exact neutral partition function (\ref{Zneuthypergeom}). 

  Before we use (\ref{Znexpanded})  to find the high-temperature asymptotics, in the next section we find it useful to compare with the expansion derived via the more physical approach of passing from the grand-canonical to the canonical partition function.

  \section{The canonical partition function} \label{Canonical}
  
In the previous sections we built the neutral partition function from the bottom up, by imposing neutrality as an exact condition and identifying the resulting $q$-series first as the non-negative crank partition function and then as the $q$-hypergeometric function (\ref{Zneuthypergeom}). Using its quantum modular properties, we expanded it at high temperature and found the expression (\ref{Znexpanded}). In this section we instead start from the full partition function (\ref{Zfulltheta}), which is in the grand canonical ensemble, and pass to the canonical ensemble to obtain the neutral sector. This is essentially the approach followed in \cite{Kraus:2011ds} in the more general case with spin-3 charges present. We will see that the partition function we obtain this way, which we will call $Z_C$, agrees with the expansion above to first order in the high-temperature expansion.

Recall that in the thermodynamic limit we can approximate the canonical partition function by performing a saddle point integration of the grand canonical partition function with respect to the spin-1 chemical potential. That is,
\be
Z_C(\tau)=\int Z(u,\tau) e^{-2\pi i u Q}\diff u=\int e^{\ln Z(u,\tau)-2\pi i u Q}\diff u\;.
\ee
We want the neutral sector $Q=0$, so the saddle point is at $\frac{\p \ln Z(u,\tau)}{\p u}=2\pi iQ=0$. Using the high-temperature expression (\ref{Zutmodular}) we find
\be
\frac{1}{\pi}\frac{\p Z(u,\tau)}{\p u}=\frac{q^{-\frac{1}{12}} e^{\pi i u^2\tauh}}{\eta(\tauh)2\cos\pi u}\left(2 i u \tauh \theta_4(-u\tauh|\tauh)+\tan(\pi u)\theta_4(-u\tauh|\tauh)-\tauh\theta_4'(-u\tauh|\tauh)\right)=0\;,
\ee
which vanishes at $u=0$. So the saddle for $\tau\ra 0$ is given by $u=0$ (in agreement with the restriction of the condition of \cite{Kraus:2011ds} to $\alpha=0$, and our expectation that the charged and neutral partition functions have the same leading behaviour). Expanding around $u=0$, we are left with the Gaussian integral
\be \label{canonicalD1}
Z_C= Z_{\text{charged}}(\tau) \int e^{\half \frac{\p^2 \ln Z(u,\tau)}{\p u^2}|_{u=0} u^2}\diff u\;.
\ee
Evaluating the integral we find the prefactor
\be
Z_C=\frac{Z_{\text{charged}}(\tau)}{\sqrt{-\frac1{2\pi} \frac{\p^2 \ln Z(u,\tau)}{\p u^2}|_{u=0}}}\;,
\ee
which gives us the relation between the grand canonical and canonical (neutral) partition functions in the thermodynamic limit.

Evaluating the second derivative at $u=0$, up to exponentially small ($\qh\ra 0$) terms, we find:
  \be
\frac{1}{\pi^2}  \frac{\p^2 \ln Z(u,\tau)}{\p u^2}|_{u=0}=(2i\tauh/\pi+1/\cos(\pi u))|_{u=0}=2i\tauh/\pi+1\;.
  \ee
  So we obtain that
  \be
  Z_C(\tau)=\frac{Z_{\text{charged}}(\tau)}{\sqrt{\frac{i}{\tau}-\frac{\pi}2}}\;.
  \ee
  Expanding in small $\tau$, we find
  \be \label{Zcanonicalfinal}
  Z_C(\tau)=Z_{\text{charged}}(\tau)\sqrt{-i\tau} (1+\frac{\pi(-i\tau)}4 +\cdots)\;,
  \ee
  which, given also (\ref{Zchargedasympt}), matches the correct exponential and first subleading behaviour of the neutral partition function (\ref{Znexpanded}). We therefore have an independent confirmation of (\ref{Znexpanded}), of course within the validity of the saddle-point approximation.

\section{The asymptotic density of states} \label{Asymptotic}

In this section we would like to find the asymptotic density of states of our neutral partition function, which is the main ingredient entering the Cardy formula. 
It is instructive to first review the computation for the full (charged) partition function, in order to highlight the differences with the neutral case. This computation is standard and reviews can be found in many sources, e.g. \cite{Carlip:2005zn,Murthy:2023mbc,Loran:2010bd}.
As discussed, when $\tau\ra0$, $q\ra 1$ and it is convenient to re-express $Z_{\text{charged}}(\tau)=Z(u\!=\!0,\tau)$ in terms of $\tauh=-1/\tau$, such that $\qh\ra 0$ (see (\ref{Zutmodular})). We define a generating function $\rho(n)$ as
\be
Z_{\text{charged}}(\tau)=\sum_{n=0}^{\infty} \rho(n) q^n=\sum_{n=0}^{\infty}\rho(n) e^{2\pi i n\tau}\;.
\ee
To obtain the asymptotic behaviour of the generating function, we perform the inverse Laplace transform
\be
\rho_{\text{charged}}(n)=\int_C Z_{\text{charged}}q^{-n}\diff\tau=\half \int_C e^{\frac{\pi i}{12\tau}}e^{-\frac{\pi i \tau}{6}}e^{-2\pi i n \tau} \diff \tau\;,
\ee
where we used (\ref{Zchargedasympt}), i.e. we dropped exponentially small terms. The contour $C$ is taken around the origin. At this point one often applies the saddle point approximation, which gives $\tau_0=i/\sqrt{24 (n+1/12)}$. This gives the correct leading expression as well as the overall $n$-dependent prefactor, however to obtain the correct subleading $1/n$ behaviour one needs to do the integral exactly. Recalling the definition of the Bessel $J$ functions
  \be
  J_n(z)=\frac1{2\pi i}\left(\frac{z}{2}\right)^n\int_C t^{-n-1} e^{t-\frac{z^2}{4t}}  \diff t\;,
  \ee
  we evaluate
  \be
  \rho_{\text{charged}}(n)=\half \int_C e^{t+1/t(\pi^2(n+1/12)/6)} \frac{\diff{t}}{-2\pi i (n+\frac{1}{12})}=\frac{i}{2\sqrt{6}}\frac{\pi}{\sqrt{n+\frac{1}{12}}} J_1 \left(\sqrt{\frac{2}{3}}\pi i \sqrt{n+\frac{1}{12}}\right)\;,
  \ee
  where we used that $J_{-n}(z)=(-1)^n J_n(z)$. Expanding the Bessel function for large $n$ (see appendix \ref{appendixA}), we obtain   
        \be \label{Zfasympt}
        \rho_{\text{charged}}(n)= \frac{e^{2\pi\sqrt{n/6}}}{ 4\cdot 6^{1/4} \cdot n^{3/4}} \left(1 \!+\! \left(\frac{\pi}{12\sqrt{6}} \!-\! \frac{3\sqrt{3/2}}{8\pi}\right)\frac{1}{\sqrt{n}} \!+\! \left(\frac{\pi^2}{1728}\!-\! \frac{45}{256\pi^2} \!-\! \frac{5}{64}\right)\frac{1}{n} \!+O\left(\frac{1}{n^{3/2}}\right)\right)
        \ee
which fully agrees with the leading and subleading terms listed in \cite{fullpartition}.

  We are now ready to find the asymptotic generating function for the neutral case. Given the relation (\ref{Zcanonicalfinal}), to see the leading order behaviour it's
  enough to simply multiply the $\rho_{\text{charged}}(n)$ we found above with $\sqrt{-i\tau_0}$ where $\tau_0=i/\sqrt{24n}$ is the (leading part of the) saddle-point value. This gives
  \be \label{rhoneutralsaddle}
  \rho_{\text{neutral}}(n)=\frac{\rho(n)}{(24 n)^{1/4}}=\frac{e^{2\pi\sqrt{n/6}}}{ 4\cdot 6^{1/4} \cdot n^{3/4}}\times\frac{1}{(24n)^{1/4}}=
\frac{e^{2\pi\sqrt{n/6}}}{8\sqrt3 n}\;,
  \ee
  which agrees with the asymptotic answer stated in \cite{Nonnegativecrank}. So the saddle point technique correctly reproduces the leading terms in the asymptotic expansion. It is perhaps worth noting that the prefactor is just half of the prefactor of the standard Hardy-Ramanujan partition sum $\sum_{n} p(n)q^n =1/(q;q)_{\infty}\sim \frac{1}{4\sqrt{3}n}e^{2\pi\sqrt{\frac{n}{6}}}$, which would be the correct asymptotic behaviour of the neutral sector of the complex fermion partition function in the NS (half-integer moded) sector (see e.g. \cite{DiFrancesco} p.391). 

\begin{figure}[ht]
  \begin{center}
\begin{tikzpicture}
  \node at (1,2) {\includegraphics[width=12cm]{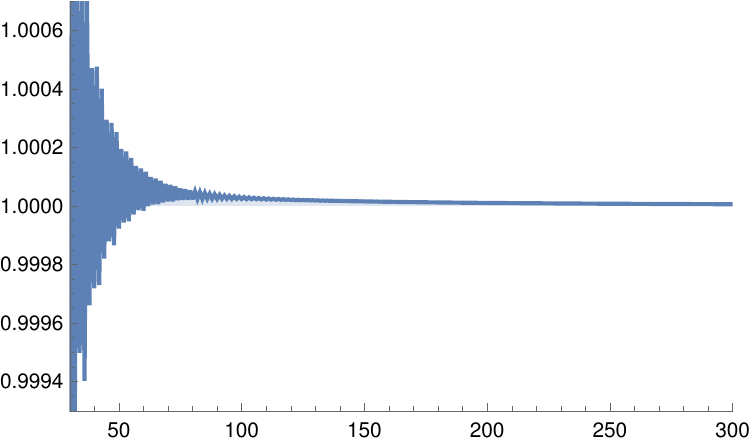}};
  \node at (2,-1.5) {$n$};
  \node at (-2.5,5.2) {$\frac{\rho_{\text{neutral}}(n)}{a(n)}$};
 \end{tikzpicture}
  \end{center}
  \caption{The ratio of $\rho_{\text{neutral}}(n)$ as calculated from the asymptotic formula (\ref{finalrho}) to the actual coefficients of the non-negative crank series as calculated from (\ref{ZUncu}), as a function of $n$. Note that even for $n\sim 40$ the functions agree to within 0.01\%.} \label{ratio}
  \end{figure}

After this initial check, let us now use the $\tau$-expansion of the neutral partition function (\ref{Znexpanded}) to go beyond the leading asymptotics. In particular, we will compute the terms which contribute to the leading $1/n$ prefactor that we saw above, as well as the $n^{-3/2}$ and $n^{-2}$ subleading behaviour. From the leading term in (\ref{Znexpanded}) we find:
  \be\begin{split}
  \rho_{\text{neutral}}^{(0)}(n)&=\half \int_C \sqrt{-i\tau} e^{-2\pi i \tau (n+1/12)+\frac{\pi}{12\tau}} \diff \tau\\
  &=  \frac{1}{2(-i)(2\pi)^{3/2}(n+1/12)^{3/2}}\int t^\half e^{t-\frac{1}{t}\frac{z^2}4}\diff t\;,
\end{split}
  \ee
    where we have again converted $t=-2\pi i (n+1/12)\tau$ and defined $z=\sqrt{2/3}\pi i \sqrt{n+1/12}$. 
    Using again the definition of the Bessel function, a straightforward computation gives\footnote{Of course $J_{3/2}$ is an elementary function,
    $J_{3/2}(z)=\sqrt{2/(\pi z)}(-\cos(z)+\sin(z)/z)$, and the same holds for the other half-integral Bessel functions. Still, we find it more insightful to organise the results in Bessel form by analogy to the integer-order case.}
    \be
    \rho_{\text{neutral}}^{(0)}=\frac{\pi}{2\sqrt{2}\cdot 6^{3/4}(n+1/12)^{3/4}} \left(-i^\half J_{3/2}(z)\right)\;.
    \ee
    Expanding the Bessel function for large $n$ (see Appendix \ref{appendixA}), we find:
  \be
  \rho_{\text{neutral}}^{(0)}(n)=\frac{ e^{\sqrt{\frac23}\pi\sqrt{n+1/12}}}{8\sqrt{3}(n+1/12)}\left(1-\frac{1}{\pi\sqrt{2/3(n+1/12)}}+O\left(\frac1{n^{3/2}}\right)\right),
  \ee
  of course in full agreement with the leading-order computation above. Let us now look at the next term in the expansion (\ref{Znexpanded}),
  \be
  \begin{split}
  \rho_{\text{neutral}}^{(1)}(n)&=\frac{\pi}{8}\int_C(-i\tau)^{3/2}e^{-2\pi i \tau(n+1/12)+\frac{\pi i}{12\tau}}\diff \tau\\
  &=\frac{\pi}{-2\pi i 8 (2\pi)^{3/2}(n+1/12)^{5/2}} \int_C t^{\frac{3}{2}}e^{t-\frac{1}{t}\frac{z^2}4}\diff t\\
  &=\frac{2\pi^2}{32\sqrt{2}\cdot 6^{5/4}(n+1/12)^{5/4}} \left(i^{\frac32} J_{5/2}(z)\right)\;.
 \end{split}
  \ee
  Expanding this expression for large $n$, we obtain
  \be
  \rho_{\text{neutral}}^{(1)}(n)=\frac{\pi e^{\sqrt{\frac23}\pi\sqrt{n+1/12}} }{16\cdot 4 \cdot 3 \cdot \sqrt{2} (n+1/12)^{3/2}}\left(1-\frac{3}{\sqrt{2/3}\pi\sqrt{n+1/12}} + O\left(\frac1{n}\right)\right)\;.
    \ee
The next term is
\be\begin{split}
\rho_{\text{neutral}}^{(2)}(n)&=\frac{5\pi^2}{64}\int_C(-i\tau)^{5/2}e^{-2\pi i \tau(n+1/12)+\frac{\pi i}{12\tau}}\diff \tau\\
&=\frac{5\pi^2}{64(-2\pi i)(2\pi)^{5/2}(n+1/12)^{7/2}}\int_C t^{5/2}e^{t-\frac{1}{t}\frac{z^2}{4}}\diff t\\
&=\frac{5\pi^3}{64\cdot 4\sqrt{2} \cdot 6^{7/4} (n+1/12)^{7/4}}\left(i^\half J_{7/2}(z)\right)\;,
\end{split}
  \ee
  and gives for large $n$ the contribution 
  \be
  \rho_{\text{neutral}}^{(2)}(n)=\frac{5\pi^2 e^{\sqrt{\frac23}\pi\sqrt{n+1/12}} }{64\cdot 32\cdot 3^{3/2} (n+1/12)^{2}}\left(1+O\left(\frac1{n^{1/2}}\right)\right)\;.
  \ee
  These are all the terms that contribute to order $1/n$ beyond the leading order. Adding them and expanding for large $n$,  we finally find the asymptotic expression
  \be \label{finalrho}
  \rho_{\text{neutral}}(n)=\frac{e^{2\pi\sqrt{\frac{n}{6}}}}{8\sqrt{3}n}\left(1+\left(\frac{5\pi}{24\sqrt{6}}-\frac{\sqrt{3}}{\sqrt{2}\pi}\right)\frac{1}{\sqrt{n}}+\left(\frac{61\pi^2}{6912}-\frac{5}{16}\right)\frac{1}{n}+O\left(\frac{1}{n^{3/2}}\right)\right)\;.
    \ee
    As can be seen in Fig. \ref{ratio}, this expression reproduces the actual coefficients $a(n)$, computed through (\ref{ZUncu}), extremely well, even for not-particularly-large values of $n$.    
It is worth mentioning that, due to the $n^{-3/4}$ prefactor, the $\rho_{\text{charged}}(n)$ asymptotics are in agreement with the logarithmic corrections to the (non-higher-spin) 3d gravity BTZ entropy, $\ln(\rho_{\text{charged}}(n))=S_0-\frac{3}2 \ln S_0 + \cdots$ \cite{Carlip:2000nv}, while the $\rho_{\text{neutral}}(n)$ asymptotics would lead to a factor or $-2$ rather than $-3/2$. The reason for the agreement is presumably that the $\rho_{\text{charged}}$ answer depends on standard modular properties, which lead to universal results. It would be very interesting to compute the logarithmic corrections for higher-spin gravity from the bulk side (which, as far as we are aware, has not been attempted, but should be achievable even at general $\lambda$) and see whether one obtains something closer to $\rho_{\text{neutral}}$.

\section{Discussion}

In this note we studied various aspects of the partition function for a neutral free complex fermion. We showed that although this partition function is not a modular form, it does have a quantum modular property in the sense of \cite{ZagierQuantum}, which allowed us to expand it in the high temperature limit and evaluate its asymptotics as well as the subleading terms. The results for the asymptotic density of states agree with the literature (as summarised in \cite{Nonnegativecrank}), and we were also able to exhibit the subleading terms which were not previously recorded.

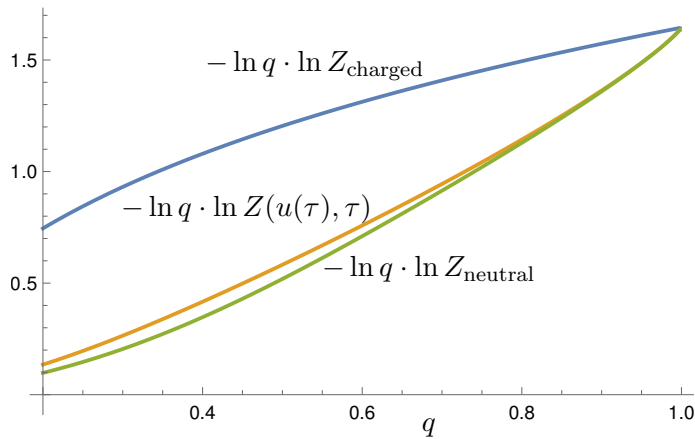
\begin{figure}[ht]
  \begin{center}
\begin{tikzpicture}[baseline={(0,0cm)}]
  \node at (1,2) {\includegraphics[width=9cm]{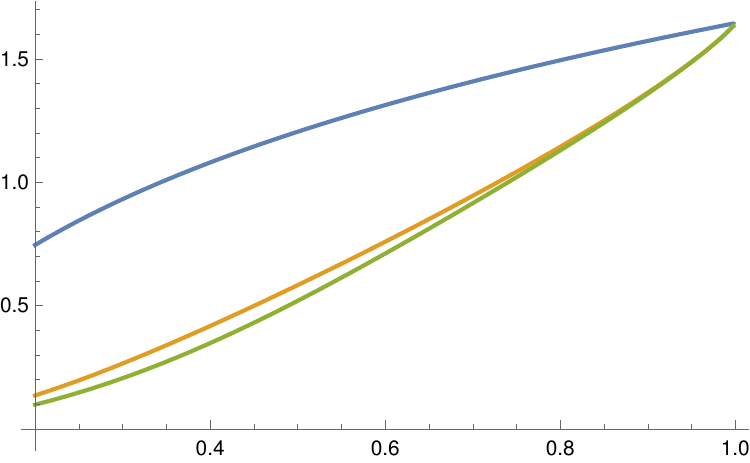}};
  \node at (2,-0.8) {$q$};
  \node at (0.5,4){$\footnotesize -\ln q\cdot \ln Z_{\text{charged}}$};
  \node at (-0.4,2.1){$\footnotesize -\ln q\cdot \ln Z(u(\tau),\tau)$};
   \node at (2,1.3){$\footnotesize -\ln q \cdot \ln Z_{\text{neutral}}$};
\end{tikzpicture}
  \end{center}
  \caption{As in Fig. 1, we plot $-\ln q\cdot \ln Z$ for the charged $u=0$ partition function (top), the neutral partition function (bottom) and now also $Z(u(\tau),\tau)$ (middle) where $u(\tau)$ is given by the first correction away from $\tau=0$ given in (\ref{ucorrection}). We see good agreement with the neutral partition function as $q\ra 1$.} \label{utau}
\end{figure}

What is perhaps more interesting is the implication that there should be a relation $u(\tau)$ which inverts the equation
\be
Z(u(\tau),\tau)=Z_{\text{neutral}}(\tau)\;,
\ee
i.e. a function which, when substituted in the full partition function leads to the neutral partition function. The reason this is interesting is that this function might be expected to have modular properties (at least perturbatively as $\tau\ra i0^+$), since both $Z(u,\tau)$ and $Z_{\text{neutral}}$ do. At low temperature, $q\ra 0$, we can solve this equation perturbatively to obtain the (not-so-enlightening) expression
\be
u(q)=-\frac{1}{4}+\frac{1}{2\pi}\left(q^2+q^4-q^5-\frac56 q^6+q^7-\frac{9}{2}q^8+\frac{17}{2}q^9+\cdots\right)\;,
\ee
while at high temperature, comparing (\ref{Zutmodular}) to (\ref{Znexpanded}) we find the (more promising) small $\tau$ behaviour
\be \label{ucorrection}
u(\tau)=\sqrt{\frac{i\tau}{2\pi}\ln(-i\tau)}+\cdots
\ee
Substituting this expression in $Z(u(\tau),\tau)$ we indeed find good agreement with the neutral partition function, as illustrated in Fig. \ref{utau}.

Recalling our motivation from section \ref{Motivation}, one is of course interested in how our results might apply to the higher-spin/$\Wcal$ algebra correspondence. A first aspect to note is that for a comparison with the bulk one needs $D$ fermions, and although in that case the grand canonical partition function is just $Z(u,\tau)^D$, the restriction to the $\zeta^0$ part as in (\ref{Zrestricted}) is not immediate. In Appendix \ref{appendixC} we work it out for $D=2$, but ideally we would require an expression for large $D$ to make contact with the bulk. The generalisation, at $D=1$, of the non-negative crank partition function to include the spin-3 chemical potential, and which has the same relation to the partition function (\ref{HSPartition}) as $Z_{\text{neutral}}$ does to (\ref{Zfull}), appears straightforward. However, one would then need to establish a suitable generalisation of the modular transformation (\ref{Hmodular}) in order to make progress.

 Of course, as a first step, one could work in the $\alpha^2$ expansion. The authors of \cite{Gaberdiel:2012yb} showed, using general considerations regarding the modular/high-temperature behaviour of traces with higher-spin field insertions, that the leading behaviour of the  $O(\alpha^{2n})$ terms of the partition function (\ref{HSPartition}) can be obtained, for general $\lambda$, and matched with the expression found from the bulk in \cite{Kraus:2011ds}. This indicates that each term in the $\alpha^2$-expansion, restricted to the $\zeta^0$ sector, should have analogous modular properties to our $Z_{\text{neutral}}$ which should allow a computation, at $\lambda=0$ and $D=1$, of the subleading (perturbative in $\tau$) corrections to the results of \cite{Kraus:2011ds,Gaberdiel:2012yb}. We hope to report on such a computation in the future. 

A recent line of study \cite{Downing:2021mfw,Downing:2023lop,Downing:2023lnp} considers the modular properties of Generalised Gibbs Ensembles for real ($c=\half$) free fermions with insertions of higher KdV charges, which are what would correspond to the higher $\Wcal$ charges in our complex fermion case. Although the details are different, one would expect that the techniques used in those works, as well as their physical interpretation of the modular transformed equations, could provide useful input in the higher-spin context as well. 
 
To conclude, we hope that our results will provide some guidance in terms of what types of modular objects one would expect to encounter in the computation of subleading corrections to the entropy in the $\Wcal_\infty[0]$ case including higher-spin chemical potentials, and their eventual comparison with bulk computations. It seems clear that, in this exact setting, going beyond saddle-point computations is possible, and also worthwhile, as it allows us to identify more precisely the types of modular properties that are behind the matching of the bulk and boundary entropy, and might lead to a more detailed understanding of the microscopic degrees of freedom.

\mbox{}

{\bf Note added in v2}: We have been informed \cite{NazarogluPrivateCommunication} that, for the $D=1$ case, it is possible to go beyond our asymptotic formula (\ref{finalrho}) and obtain a \emph{convergent} formula for the coefficients of $Z_{\text{neutral}}$. More specifically, with the numerator of $Z_{\text{neutral}}$ appropriately understood as a unary false theta function, and given that the overall weight of $Z_{\text{neutral}}$ is zero, the techniques developed in \cite{Bringmann:2019vyd} allow the computation of an exact Rademacher-type formula. This can be truncated to a finite series for numerical evaluation, with the precision increasing with the number of terms included. The resulting series expressions can be seen to agree with the exact coefficients $a(n)$ to far greater accuracy than the asymptotic formula, even for very low values of $n$. As first discussed in the IIB string theory context in \cite{Dijkgraaf:2000fq} (see \cite{Murthy:2023mbc} for a review and further references), having a full Rademacher-type expansion instead of just an asymptotic one provides a much more detailed understanding of the subleading bulk contributions to the gravitational path integral. One could therefore hope for a similar understanding to eventually emerge in our higher-spin gravity context as well. For that it would be critical to understand whether similar exact expressions are possible for the higher-$D$ cases, for which perhaps the extensions of \cite{Bringmann:2019vyd} to higher rank and depth \cite{Bringmann:2021wuk, Bringmann:2021dxg} will be of relevance.  We intend to come back to these questions in future work.

\paragraph{Acknowledgements} We are thankful to Sam van Leuven for very enlightening discussions, and to Caner Nazaroglu for very helpful and far-reaching communications regarding our work.  
MM was supported by a DST/NRF Innovation Postdoctoral Fellowship during the initial stages of this work, and by the National Institute for Theoretical and Computational Sciences (NITheCS) during the later stages.

\appendix

\section{Definitions} \label{appendixA}

In this appendix we collect some standard definitions and results. In terms of $q=e^{2\pi i\tau}$ and $\zeta=e^{2\pi i u}$, the  product and series forms of the Jacobi theta functions are given by:
\be
\theta_1(u|\tau)=2q^{\frac18}\sin(\pi u)\prod_{n=1}^{\infty}(1-q^n)(1-q^n\zeta)(1-q^n/\zeta) = -i \sum_{r\in \Zset+\half}(-1)^{r-\half} \zeta^r q^{r^2/2}\;,
\ee

\be
\theta_2(u|\tau)=2q^{\frac18}\cos(\pi u)\prod_{n=1}^{\infty}(1-q^n)(1+ q^n\zeta)(1+q^n/\zeta) = \sum_{r\in \Zset+\half} \zeta^r q^{r^2/2}\;,
\ee

\be
\theta_3(u|\tau)=\prod_{n=1}^{\infty}(1-q^n)(1+ q^{n-\half}\zeta)(1+q^{n-\half}/\zeta) = \sum_{n\in \Zset} \zeta^n q^{n^2/2}\;,
\ee
and 
\be
\theta_4(u|\tau)=\prod_{n=1}^{\infty}(1-q^n)(1- q^{n-\half}\zeta)(1-q^{n-\half}/\zeta)=\sum_{n\in \Zset}(-1)^n \zeta^n q^{n^2/2}\;,
\ee
while we also have
\be
\theta_1'(u|\tau)=\frac{\p}{\p u}\theta_1(u|\tau)=2\pi q^{\frac18}\sum_{k=0}(-1)^k(2k+1)\cos(\pi(2k+1)u) q^{\frac{k(k+1)}{2}} \;.
\ee
We also recall the Dedekind eta function
\be
\eta(\tau)=q^{\frac{1}{24}}\prod_{n=1}^\infty (1-q^n)\;,
\ee
and the $q$-Pochhammer symbol, defined as
\be
(a;q)_n=(1-a)(1-aq)\cdots (1-aq^{n-1})\;,
\ee
so that $\eta(\tau)=q^{\frac1{24}}(q;q)_\infty$.

The $\theta$ and $\eta$ functions are modular of weight $1/2$. The transformations we will need are:
\be
\theta_4(u/\tau|-1/\tau)=\sqrt{-i\tau} e^{\pi u^2/\tau}\theta_2(\tau)\quad\text{and}\quad
\eta(-1/\tau)=\sqrt{-i\tau}\eta(\tau)\;.
\ee
$\theta_1'$ is also modular, of weight $3/2$. We will only need $\theta_1'(0,\tau)$, whose modular transformation follows from the equality
\be
\theta_1'(0,\tau)=2\pi \left(\eta(\tau)\right)^3\;.
\ee
The $q$-hypergeometric function as defined in \cite{Folsometal13} is
\be
F(a,b,t;q)=\sum_{k=0}^\infty \frac{(aq;q)_k}{(bq;q)_k} t^k\;.
\ee
However, a more standard definition of the $q$-hypergeometric series (as also implemented in Mathematica) is:
\be
{}_r\phi_s(a_1\ldots a_r;b_1\ldots b_s;q;z)=\sum_{k=0}^\infty \frac{(a_1;q)_k \cdots (a_r;q)_k}{(b_1;q)_k\cdots (b_s;q)_k} \left((-1)^k q^{k(k-1)/2})\right)^{1+s-r} \frac{z^k}{(q;q)_k}\;,
\ee
where we note the additional $(q;q)_k$ in the denominator. In terms of this definition, the function we are interested in can be expressed as
  \be
  q^{-\frac18}H(1,2;\tau)= F(-q^{-1},-1;-q;q^2)=\sum_{k=0}^\infty \frac{(-q,q^2)_k}{(-q^2,q^1)}(-q)^k={}_2\phi_1(q^2,-q;-q^2;q^2,-q)\;.
\ee 

We will also need the asymptotic expansions of the $J_k$  Bessel functions (see e.g. \cite{AbramowitzStegun}):
\be
J_k(z)=\frac{e^z}{\sqrt{2\pi z}}\left(1-\frac{4k^2-1}{8z}+\frac{(4k^2-1)(4k^2-9)}{2!(8z^2)}+ O(1/z^3)\right)\;.
\ee
To let the argument take large imaginary values we use the analytic continuation $J_k(iz)=e^{\frac{k\pi i}{2}} J_k(z)$, 
to find, as $z\ra i \infty$, 
\be
-i J_1(z)=\frac{e^{-iz}}{\sqrt{2\pi(-iz)}}\left(1-\frac{3}{8(-iz)}-\frac{15}{128 (-iz)^2}+O(1/z^3)\right)\;,
\ee
\be
-i^\half J_{3/2}(z)=\frac{e^{-iz}}{\sqrt{2\pi (-iz)}}\left(1-\frac1{(-i z)}+O(1/z^3)\right)\;,
\ee

\be
 i^{\frac32} J_{5/2}(z)=\frac{e^{-iz}}{\sqrt{2\pi (-iz)}}\left(1-\frac{3}{(-iz)}+\frac{3}{(-iz)^2}+O(1/z^3)\right)\;,
 \ee
and
 \be
 i^{\frac12} J_{7/2}(z)=\frac{e^{-iz}}{\sqrt{2\pi (-iz)}}\left(1-\frac{6}{(-iz)}+\frac{15}{(-iz)^2}+O(1/z^3)\right)\;,
\ee
where the factors of $i$ in front are such that the right-hand sides are real and positive. Of course, all the half-integral order cases are expressible in terms of elementary functions, so one could have found the asymptotics in a simpler way as well, but keeping them as Bessel functions helps with the organisation of the results.

\section{Free fermions} \label{appendixB}

For completeness, in this appendix we record the relevant generators of $\Wcal_{1+\infty}$ \cite{Pope:1990kc}, in its fermionic realisation (see e.g. \cite{Pope:1991ig} for a review of the various realisations of $\Wcal_{1+\infty}$ and \cite{Awata:1993vn} for more details on the computation of the character formula). More details can be found in \cite{Kraus:2011ds}, from whose conventions we differ just by simple rescalings to simplify the presentation. We consider a complex fermion with OPE
\be
\psibar(z)\psi(w)\sim \frac{1}{z-w}\;,
\ee
and expand it in modes as $\psi(z)=-i^\half \sum_{n\in \Zset} b_n e^{inz}$, $\psibar(z)=-i^\half \sum_{n\in \Zset} \bbar_n e^{inz}$, where $z$ is the cylinder coordinate, identified as $z\sim z+2\pi$. Since $n$ is integer, this mode expansion leads to periodic boundary conditions on the cylinder, which is the case relevant to excitations around the BTZ vacuum \cite{Coussaert:1993jp}. The modes satisfy $\{\bbar_m,b_n\}=\delta_{m,-n}$, and the spin 1,2, and 3 generators are:
\be
\begin{split}
\Qcal&=-\sum_{n\geq 1} \left(\bbar_{-n} b_n-b_{-n}\bbar_n\right)\;,\\
\Lcal&=-\sum_{n\geq 1} n \left(\bbar_{-n} b_n+b_{-n}\bbar_n\right)\;,\\
\Wcal&=-\sum_{n\geq 1} n^2\left(\bbar_{-n} b_n-b_{-n}\bbar_n\right)\;.\\
\end{split}
\ee
where in $\Lcal$ we have subtracted the $1/24$ zero-point contribution so that our $q$-series start with 1. We emphasise that, compared to \cite{Kraus:2011ds} we have rescaled $\Wcal$ and absorbed various numerical factors into the chemical potential $\alpha$ appearing in (\ref{HSPartition}).

Defining the vacuum to be annihilated by the positive modes, $b_{n}\ket{0}=\bbar_n\ket{0}=0$ for $n>0$, and creating states by acting with all possible $b_{-n}$, $\bbar_{-n}$ on the vacuum, one obtains the charges listed in Table \ref{states}, and eventually (\ref{HSPartition}).

 Note that we are only considering the left-movers, as the right-movers can be treated similarly.

\section{The $D=2$ case}  \label{appendixC}

In section \ref{Neutral} we saw how the restriction of $Z(u,\tau)$ to the neutral sector works for a single fermion, $D=1$. To get a flavour of the complications that will arise in the general case, in this appendix we derive the corresponding result for $D=2$.

For two complex fermions, the grand-canonical partition function is just the square of (\ref{Zfull}). Expanding and keeping the $\zeta^0$ terms, we find that the neutral-sector series goes like\footnote{It should be emphasised that, unlike the gauge-theory-motivated setting of e.g. \cite{Razamat:2012uv,Pan:2021mrw}, we still have only one $\Urm(1)$ current, defined as $\sum_i\psibar^i\psi^i$, $i=1,2$, so even though $D=2$ we are still only integrating over one $\zeta$.}  
\be \label{neutralD2exp}
Z_{\text{neutral}}^{D=2}=\oint \frac{\diff \zeta}{2\pi i\zeta}(Z(u,\tau))^2=1+4q^2+8q^3+13q^4+24q^5+46q^6+80q^7+139q^9+\cdots
\ee
To find the all-orders formula, we start by recalling that
\be
\begin{split}
\theta_2^2(u|\tau)&=\half\left(\theta_3(0|\tau/2)\theta_3(u|\tau/2)-\theta_4(0|\tau/2)\theta_4(u|\tau/2)\right)\\
&=\half\left(\sum_{k\in \Zset} q^{\frac{k^2}{4}}\sum_{\ell\in\Zset}\zeta^\ell q^{\frac{\ell^2}{4}}-
\sum_{k\in \Zset} (-1)^k q^{\frac{k^2}{4}}\sum_{\ell\in\Zset}(-1)^\ell \zeta^\ell q^{\frac{\ell^2}{4}}\right)\\
&=\half\sum_{k\in \Zset}\sum_{\ell\in\Zset}\left(1-(-1)^{k+\ell}\right)\zeta^\ell q^{\frac{k^2+\ell^2}{4}}\\
&=\sum_{k,\ell \in \Zset,k+\ell \;\text{odd}} \zeta^\ell q^{\frac{k^2+\ell^2}{4}}\;.
\end{split}
\ee
As indicated, the nonzero terms are those for which $k+\ell$ is odd. Considering the contributions to a fixed power of $q^{\frac{k^2+\ell^2}4}$, we can distinguish two cases:

\begin{itemize}
\item If $k=0$ and $|\ell|=n>0$, we get $\zeta^n+\zeta^{-n}$, while the contribution from $\ell=0$ and $|k|=n$ gives 2. Adding these gives $(2+\zeta^n+\zeta^{-n})q^{n^2/4}$.

\item If $|k|=m>0$ and $|\ell|=n>m$, then the contribution will be $2(\zeta^{n}+\zeta^{-n})q^{\frac{m^2+n^2}4}$ (since both $k$ and $-k$ contribute equally), and the contribution from $|k|=n$ and $|\ell|=m$ will give $2(\zeta^{m}+\zeta^{-m})q^{\frac{m^2+n^2}4}$. So in total we get a term $2(\zeta^{m}+\zeta^{-m}+\zeta^{n}+\zeta^{-n})q^{\frac{m^2+n^2}4}$. 
\end{itemize}
So the series becomes
  \be
\theta_2^2(u|\tau)=\sum_{n>0, n \;\text{odd}} (2+\zeta^n+\zeta^{-n})q^{\frac{n^2}4}+2\sum_{\substack{m>0, n>m\\m+n \;\text{odd}}}(\zeta^{m}+\zeta^{-m}+\zeta^{n}+\zeta^{-n})q^{\frac{m^2+n^2}4}\;.
\ee
Now we need to divide by  $1/(2\cos(\pi u))^2=1/(\zeta^\half+\zeta^{-\half})^2$. To extract the power of $\zeta^0$, we write the generic term (including  $m=0$ as a special case) as
\be
\begin{split}
  \quad\frac{\zeta^{m}+\zeta^{-m}+\zeta^{n}+\zeta^{-n}}{(\zeta^\half+\zeta^{-\half})^2}&=
  \frac{\zeta^{\frac{n-m}2}+\zeta^{-\frac{n-m}2}}{\zeta^\half+\zeta^{-\half}}\cdot\frac{\zeta^{\frac{m+n}2}+\zeta^{-\frac{m+n}2}}{\zeta^\half+\zeta^{-\half}}\\
  &=\left(\zeta^{-n+m+1}\!-\!\zeta^{-n+m+3}+\cdots -\zeta^{n-m-3}+\zeta^{n-m-1}\right) \\
  &\qquad \cdot \left(\zeta^{-m-n+1}\!-\!\zeta^{-m-n+3}+\cdots -\zeta^{m+n-3}+\zeta^{m+n-1}\right)\;.
\end{split}
\ee
The first parenthesis has $n\!-\!m$ terms, and each one of them will multiply with a term from the second parenthesis to give a $\zeta^0$. One can check that these terms will come with a sign of $(-1)^{\frac{m-n-1}2}(-1)^{\frac{m+n-1}2}=-(-1)^m$. So we finally find that
\be\label{D2exp}
\begin{split}
  Z_{\text{neutral}}^{D=2}&=\frac{q^{-\frac14}}{(q;q)_{\infty}^2}\oint \frac{\diff \zeta}{2\pi i \zeta}\left(\frac{\theta_2(u|\tau)}{2\cos\pi u}\right)^2
  =\frac{1}{(q;q)_{\infty}^2}\sum_{\substack{m\geq 0,n>m\\m+n\; \text{odd}}}(2-\delta_{m,0})(-1)^m(n-m)q^{\frac{m^2+n^2-1}4}\\
  &=\frac{1}{(q;q)_{\infty}^2}\left(1\!-\!2q\!+\!3q^2\!+\!2q^3\!-\!6q^4\!+\!3q^6\!+\!6q^7\!-\!10q^9\!+\!2q^{10}\!-\!6q^{11}\!+\!7q^{12}\!+\!10q^{13}+\cdots\right)\;.
\end{split}
\ee
One notices that, for our constraints $m+n$ odd, $n>m$, the combinations $\frac{m^2+n^2-1}4$ reduce to $a+b$, where $a,b$ are the triagonal numbers $0,1,3,6,10,15,\ldots$. We note that there are degeneracies in the exponents, e.g. $6=0+6=3+3$, and it is the sum of the relevant contributions which appears in the series. 

This slightly peculiar $q$-series agrees, after expanding the prefactor, with the  empirical result (\ref{neutralD2exp}). Although in this case we are not aware of any mathematical results regarding its modular properties, the physical setting is precisely the same as the $D=1$ case, so we can expect that (a) its leading behaviour for $\tau\ra i0^+$ will be the same as the unconstrained partition function, i.e. $\sim e^{\frac{\pi i}{6\tau}-\frac{\pi i\tau}3}$ (just the square of the $D=1$ case), and more importantly (b) that $\hat{H}(\tau):=(q;q)_{\infty}^2 Z_{\text{neutral}}^{D=2}$ has a well-defined expansion perturbatively in $\tau$, analogous to (\ref{Hexpansion}). If we further assume that, as for $D=1$, the restriction to the neutral sector does not change the weight\footnote{This assumption is clearly not true in general, for instance restricting $\theta_3(u|\tau)$ to the neutral sector one would just get a constant. It is motivated by the physical expectation that the $D=2$ neutral CFT partition function would have weight zero overall, as for $D=1$. The expected small-$\tau$ expansion of $\hat{H}$   does not rely on this assumption, as it would come from the anomalous modular term.}, this would lead to a quantum modular form of weight 1. Of course, the first statement more or less follows from the $1/(q;q)_\infty^2$ factors (expressed in terms of $\eta(\tau)^2$ and modular transformed), as the remainder is unlikely to alter the exponential behaviour. Evaluating the series for large $n$, we indeed find very good numerical agreement with the leading behaviour
\be \label{D2asympt}
\rho_{\text{neutral}}^{D=2}(n)\sim\frac{e^{\sqrt{\frac{2}{3}}\pi\sqrt{2(n+1/6)}}}{16\sqrt{3}(n+1/6)}\;.
\ee
Repeating the saddle-point computation (\ref{canonicalD1}), where $u=0$ is still the saddle-point value, we find
\be
Z_{\text{neutral}}^{D=2} = \frac{e^{\frac{\pi i}{6\tau}-\frac{\pi i\tau}{3}}}{4\sqrt{2}}\sqrt{-i\tau}\left(1+\frac{\pi(-i\tau)}4+\cdots\right)\;,
\ee
where the only essential difference is the extra $\sqrt{2}$, which, when passing to the microcanonical ensemble, will be compensated by the new saddle value $\tau_0=i/\sqrt{12(n+1/6)}$. So for the leading behaviour we can just take $n+\frac{1}{12}\ra 2n+\frac{1}{3}$ everywhere, in agreement with (\ref{D2asympt}).  However, looking at the required $\tau\ra 0$ behaviour of (\ref{D2exp}), there are two $(q;q)_\infty$ factors in the denominator which will give a $(\sqrt{-i\tau})^2$. But the saddle expansion only has one $\sqrt{-i\tau}$, so to match with (\ref{D2exp}) the $\hat{H}$ series must provide a $1/\sqrt{-i\tau}$. We also need to unabsorb the factor of $q^{-1/4}$ to give the required $e^{-\pi i\tau/3}$. So a more precise statement of (b) would be that $\hat{H} \sim \frac{q^{-1/4}}{\sqrt{-i\tau}}\left(1+\pi (-i\tau)/4+O(\tau^2)\right)$.

This example hopefully illustrates the complications that will arise in finding a combinatoric expression for general $D$, but also that it would be worthwhile doing so, since the resulting series would still be expected to have analogous modular properties to the $D=1$ case.

\bibliography{modular}
\bibliographystyle{utphys}

\end{document}